\documentclass[conference, letterpaper]{IEEEtran}
\IEEEoverridecommandlockouts
\usepackage[latin1]{inputenc}
\usepackage{graphicx}
\usepackage{setspace}
\usepackage{amsmath}
\usepackage{amsfonts}
\usepackage{amssymb}
\usepackage{subfigure}
\usepackage{stmaryrd}
\usepackage{units}
\usepackage{nicefrac}
\usepackage[printonlyused]{acronym}
\usepackage{pstricks, pst-plot, pst-grad, pstricks-add, pst-node, pstricks-add}

\usepackage{hyperref}

\usepackage{balance}
\usepackage{flushend}

\newcommand{\Set}[1]{\mathcal{#1}}

\newcommand{\Tr}[0]{\text{tr}}
\newcommand{\diag}[0]{\text{diag}}

\newcommand{\etal}{\textit{et al.}}

\newcounter{TBDCounter}
\newtheorem{theorem}{Theorem}

\newcommand{\argmax}{\arg\!\max}

\acrodef{bpcu}[bpcu]{bits per channel use}
\acrodef{CAS}[CAS]{computational aware scheduling}
\acrodef{CDF}[CDF]{cumulative distribution function}
\acrodef{CCDF}[CCDF]{complementary CDF}
\acrodef{GPP}[GPP]{general purpose processor}
\acrodef{MCS}[MCS]{modulation-and-coding scheme}
\acrodef{MRS}[MRS]{maximum rate scheduling}
\acrodef{NFV}[NFV]{Network Function Virtualization}
\acrodef{RAN}[RAN]{radio access network}
\acrodef{RAP}[RAP]{radio access point}
\acrodef{SCC}[SCC]{scheduling with complexity cut-off}
\acrodef{SNR}[SINR]{signal-to-noise ratio}
\acrodef{PPP}[PPP]{Poisson point process}
\acrodef{SINR}[SINR]{signal-to-interference-and-noise ratio}
\acrodef{BS}[BS]{base station}
\acrodef{FEC}[FEC]{forward error correction}
\acrodef{UE}[UE]{user equipment}
\acrodef{RB}[RB]{resource block}
\acrodef{CBLER}[CBLER]{code block error rate}
\acrodef{TBLER}[TBLER]{transport block error rate}
\acrodef{CB}[CB]{code block}
\acrodef{TB}[TB]{transport block}
\acrodef{SWF}[SWF]{scheduling with water-filling}

\newrgbcolor{darkgreen}{0.2 0.6 0.2}
\newrgbcolor{darkred}{0.6 0.2 0.2}
\newrgbcolor{darkblue}{0.2 0.2 0.6}

\begin{document}

\title{Computationally Aware Sum-Rate Optimal Scheduling for Centralized Radio Access Networks }


\author{
\IEEEauthorblockN{Peter Rost\IEEEauthorrefmark{1}, Andreas Maeder\IEEEauthorrefmark{1}, Matthew C. Valenti\IEEEauthorrefmark{2}, and Salvatore Talarico\IEEEauthorrefmark{2}}
\IEEEauthorblockA{%
\IEEEauthorrefmark{1}NEC Laboratories Europe, Heidelberg, Germany.\\
\IEEEauthorrefmark{2}West Virginia University, Morgantown, WV, USA.}
\thanks{The research leading to these results has received partly funding from the European Union Seventh Framework Programme (FP7/2007-2013) under grant agreement n\textordmasculine~317941 (www.ict-ijoin.eu).
  The authors would like to acknowledge the contributions of their colleagues in iJOIN, although the views expressed are those of the authors and do not necessarily represent the project.}
}

\maketitle

\begin{abstract}
  In a centralized or cloud \acl{RAN}, certain portions of the digital baseband processing of a group of several \aclp{RAP} are processed at a central data center.  Centralization improves the flexibility, scalability, and utilization of computational assets. However, the performance depends critically on how the limited data processing resources are allocated to serve the needs of the different wireless devices.  As the processing load imposed by each device depends on its allocated transmission rate and channel quality, the rate-allocation aspect of the scheduling should take into account the available computing resources.  In this paper, two computationally aware schedulers are proposed that have the objective of maximizing the system sum-rate while satisfying a constraint on the offered computational load. The first scheduler optimally allocates resources and is implemented according to a water-filling algorithm. The second scheduler is suboptimal, but uses a simpler and intuitive complexity-cut-off approach. The performance of both schedulers is evaluated using an LTE-compliant system level simulator.  It is found that both schedulers avoid outages that are caused by an overflow of computational load (i.e., computational outages) at the cost of a slight loss of sum-rate.
\end{abstract}
\IEEEpeerreviewmaketitle


\section{Introduction}
The 5\textsuperscript{th} generation of mobile communication networks will be accompanied by a paradigm shift towards virtualization and ``cloudification''.  This trend is already underway: \ac{NFV} is being developed (see, for example, use case \#6 in \cite{bib:etsi_nfv_use_cases}), while requirements studies on 5G consider network and service flexibility as one of the key requirements \cite{bib:ngmn}.
One technology that offers this flexibility in the \ac{RAN} is \emph{Cloud-RAN}, which centralizes part of the radio network functions in order to exploit centralization and coordination gains \cite{Rost.etal.ComMag.2014}.

\begin{figure}[t]
	\centering
	\includegraphics[width=0.9\linewidth]{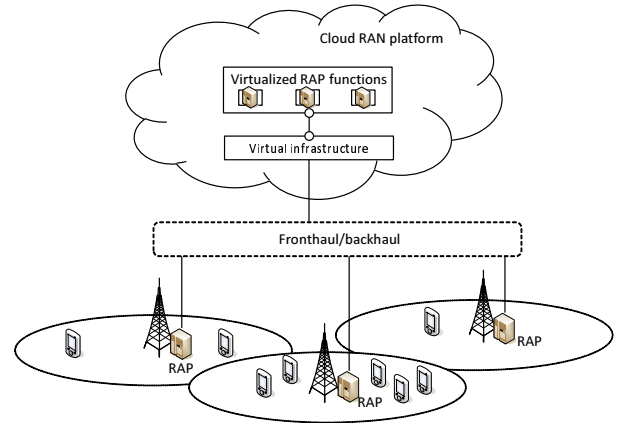}
	\caption{An exemplary Cloud-RAN architecture.}
	\label{fig:cloud_ran}
\end{figure}

In a Cloud-RAN network, the radio protocol stack is executed by different physical entities. For a given group of cells, the lower parts of the protocol stack are executed at the \acp{RAP}, while the upper parts are executed at a central entity, called the \emph{Cloud-\ac{RAN} platform}. The \acp{RAP} use dedicated hardware, while the Cloud-RAN may be implemented with commodity general-purpose hardware. Fig. \ref{fig:cloud_ran} illustrates the high-level architecture of a Cloud-RAN system. The physical \acp{RAP} are connected via a backhaul network (often also referred to as fronthaul) to the Cloud-\ac{RAN} platform, which executes the upper parts of the protocol stack (denoted as \emph{virtualized \ac{RAP} functions}). The virtual infrastructure provides on-demand computing resources, e.g. in form of processors, to the virtual \acp{RAP} \cite{Rost.etal.ComMag.2014,Wuebben.etal.SPM.2014}.  For this paper, it is important to note that the \ac{FEC}, which is amongst the most computationally  intensive \ac{RAN} functions, is executed on the Cloud-\ac{RAN} platform.

The  Cloud-\ac{RAN} platform may dynamically assign data processing resources to each \ac{RAP} based on its processing demand.  This implies the need for a mechanism which estimates the computational load of the aggregated cells and assures that the computational resources are sufficient.  In \cite{Zhu.etal.ACMCF.2011}, Zhu \etal~ demonstrate the general feasibility of this approach with a WiMAX base station implemented on \acp{GPP}. In
\cite{Bhaumik.Chandrabose.Jataprolu.Kumar.Muralidhar.Polakos.Srinivasan.Woo.MobiCom.2012}, Bhaumik \etal~ provide a numerical analysis of the required data processing resources to operate a 3GPP LTE \ac{RAN} protocol stack.  In \cite{werthmann2013multiplexing}, Werthmann \etal~present  a method based on admission control to manage the required data processing resources of a fully centralized \ac{RAN}. In \cite{Valenti.Talarico.Rost.GC.2014, Rost.Talarico.Valenti.TWC.2015}, we investigate the trade-off
between invested data processing resources and achievable data rates, taking into account specifically the processing requirements of the \ac{FEC} on the uplink.  Using this framework,
the system can be dimensioned according to the probability that the system has insufficient data processing resources  to  process  all  incoming  transmissions,  a condition that is called \emph{computational outage}.

In order to avoid a scenario where uplink transmissions are dropped due to computational outage, this paper proposes two resource-allocation strategies, which allocate resources to uplink users under a computational-complexity constraint with the goal of maximizing the overall throughput of the network.  By \emph{resource allocation} we refer to the allocation of rate to each user.  Due to the correspondence between rate and computational load \cite{Valenti.Talarico.Rost.GC.2014, Rost.Talarico.Valenti.TWC.2015}, the rate allocation problem is equivalent to allocating computational resources.  The formulation leads to a water-filling approach to the allocation.

The paper is organized as follows. Section \ref{sec:cloud_ran} introduces the system model and computational complexity framework. Section \ref{sec:rrm} derives the resource allocation strategies. Section \ref{sec:results} describes the system-level simulator and provides a comparison of two scheduling strategies and a benchmark algorithm. The paper concludes in Section \ref{sec:conclusion}.

\section{System Model}\label{sec:cloud_ran}
Consider a system with $N_c$ \acp{RAP}, which are centrally processed at the Cloud-\ac{RAN} platform. For the sake of a simplified notation, assume that each \ac{RAP} serves exactly one \ac{UE} over the complete bandwidth. This, however, does not limit the applicability of the algorithms as they operate on a per-user level and therefore can easily be extended to the case of mutiple users per \ac{RAP}.
The channel gain between \ac{RAP} and \ac{UE} remains fixed during each scheduled transmission (e.g., each \emph{subframe}), and the channel is subject to additive white Gaussian noise. The instantaneous \ac{SINR} between \ac{RAP} $k$ and its \ac{UE} for a transmission is given by $\gamma_k$. The \ac{RAP} chooses a data rate $r_k < \log(1 + \gamma_k)$ (in units of \ac{bpcu}) from a discrete set of $N_R$ available \acp{MCS}.
Let $\Set{R}_i = \left\{r_1, \dots r_{N_c}\right\}$ be the set of rates allocated to each uplink for the $i^{th}$ resource allocation strategy. Then, the set $\Set{R}^* = \left\{\Set{R}_1, \dots \Set{R}_{N_A}\right\}$ with $N_A \leq N_R^{N_c}$ denotes the set of all feasible rate allocations.


In \cite{Valenti.Talarico.Rost.GC.2014, Rost.Talarico.Valenti.TWC.2015}, the concept of \emph{computational outage} and the trade-off between \emph{computational complexity} and data rate was discussed using the model introduced by Grover \etal~in \cite{Grover.Woyach.Sahai.JSAC.2011}.  For turbo-decoding (as e.g. used in LTE), the number of turbo-iterations required to successfully decode a codeword for a particular \ac{SINR} $\gamma_k$ depends strongly on the selected data rate $r_k$. If $r_k$ is chosen close to the channel capacity $\log(1 + \gamma_k)$, on average a large number of iterations will be required. However, as the allocated $r_k$ decreases for the given \ac{SINR}, the number of required iterations also decreases. The overall computational complexity to process one codeword scales with the number of information bits that are processed and the number of iterations. The computational complexity can be computed as the product of the information bits and turbo-iterations required, divided by the number of channel uses, i.\,e. bit-iterations per channel use (pcu).

This relationship is formalized by \cite[Eq. (3)]{Rost.Talarico.Valenti.TWC.2015}
\begin{equation}
	\mathcal{C}(\gamma_k, r_k) = \frac{r_k}{\log_2\left(\zeta - 1\right)}\bigl[\log_2\left(\frac{\zeta - 2}{K(\hat\epsilon_\text{channel})\zeta}\right)
	- 2l(\gamma_k, r_k) \bigr], \label{eq:max.comp:100}
\end{equation}
where $\zeta$ is a parameter of the model related to the connectivity of the decoder when represented as a graph,
\begin{eqnarray}
K(\hat\epsilon_\text{channel}) &=&  -\frac{K'}{\log_{10}\left( \hat\epsilon_\text{channel}\right)}, \\
	l(\gamma_k, r_k) &=& \log_2\left[\log_2(1 + \gamma_k) - r_k\right],
\end{eqnarray}
$K'$ is a parameter of the model, and $\hat\epsilon_\text{channel}$ is a constraint on the channel outage probability. The set of parameters $\{ K', \zeta\}$ is selected by calibrating (\ref{eq:max.comp:100}) with an actual turbo-decoder implementation or message-passing decoder. In the remainder of this paper, we use the abbreviation $\Set{C}_k = \Set{C}(\gamma_k, r_k)$.

\section{Resource Management in Cloud-RAN}\label{sec:rrm}

This section begins in Sec. \ref{sec:Max-Rate_Optimization_Problem} with the derivation of an optimal water-filling scheduling policy when the rates are drawn from a continuum of possible rates.  Next, in Sec. \ref{sec:Optimal_Scheduler} the optimization is modified to account for the more practical case of having an \ac{MCS} scheme with only finite number of possible rates to pick from.  Finally, in Sec. \ref{sec:SCC_Scheduler}, an alternative low-complexity resource-allocation strategy is proposed that is based on an intuitive complexity cut-off approach.

\subsection{Max-Rate Optimization Problem}\label{sec:Max-Rate_Optimization_Problem}

Suppose that the Cloud-\ac{RAN} platform supports a maximum computational complexity $\Set{C}_\text{server}$, i.\,e., $\sum_k \Set{C}_k \leq \mathcal{C}_\text{server}$ must hold. Further, assume that the set of allocated rates may be drawn from a continuum of possibilities.  Our objective is to maximize the sum-rate of the system while avoiding any computational outage, which can be formalized as follows:
\begin{eqnarray}
	\Set{R}_\mathsf{opt} & = & \argmax\limits_{\Set{R} \in \Set{R}'} \sum\limits_{r_k\in\Set{R}} r_k, \label{eq:max.comp:10}\\
	& & \text{s.t.} \sum\limits_{r_k\in\Set{R}} \Set{C}_k \leq \mathcal{C}_\text{server}, \nonumber
\end{eqnarray}
where $\Set{R}'$ is the set of all rate allocations satisfying $0 \leq r_k \leq \log(1+\gamma_k)$ for all $k$.


This scheduling problem implies a trade-off between the invested computational complexity and the gained achievable rates.
In order to solve (\ref{eq:max.comp:10}), the partial derivative $\partial\mathcal{C}_k/\partial r_k$ is required, which determines how much the computational complexity varies when the data rate is changed. However, the derivative of $\mathcal{C}_k$ is rather complex and difficult to apply in this optimization problem. Therefore, we use the following piecewise linearization of $l(\gamma_k, r_k)$, which does not alter significantly the accuracy of the analysis:
\begin{eqnarray}
	l(\gamma_k, r_k) & \approx & a_k r_k + b_k,\label{eq:max.comp:20}
\end{eqnarray}
where
\begin{align*}
	a_k & = \frac{\partial l(\gamma_k, r_k)}{\partial r_k}
= \frac{-1}{\log(2)\left[\log_2\left(1 + \gamma_k\right) - r_k\right]}, \\
	b_k &  = \log_2\left[\log_2(1 + \gamma_k) - r_k\right] - a_k r_k.
\end{align*}
Using (\ref{eq:max.comp:20}), we can rewrite (\ref{eq:max.comp:100}) as
\begin{eqnarray}
	\Set{C}_k & \approx & \alpha_k r_k^2 + \beta_k r_k\label{eq:max.comp:130},
\end{eqnarray}
where
\begin{eqnarray*}
	\alpha_k & = & -\frac{2a_k}{\log_2\left(\zeta - 1\right)} \\
	\beta_k & = & \frac{1}{\log_2\left(\zeta - 1\right)}\left[\log_2\left(\frac{\zeta - 2}{K(\hat\epsilon_\text{channel})\zeta}\right) - 2b_k\right].
\end{eqnarray*}

\begin{theorem}\label{theorem:max_rate_scheduling}
	The solution to the rate allocation problem in (\ref{eq:max.comp:10}) can be well approximated by
	\begin{equation}
		r_k = \frac{1}{2\alpha_k}\left(\frac{1}{\eta} - \beta_k\right)^+, \label{eq:max.comp:135}
	\end{equation}
	where $0 \leq \eta \leq 1/\beta_k$ and $\sum\limits_{r_k\in\Set{R}} \Set{C}_k \leq \mathcal{C}_\text{server}$.
\end{theorem}

\begin{IEEEproof}
	The proof follows similar arguments and methodology of the power allocation problem \cite{Wei.Cioffi.ICC.2001}, which leads to the well known {\em water-filling} method. The details of the derivation are provided in Appendix \ref{sec:proof.theorem:max_rate_scheduling}.
\end{IEEEproof}

In (\ref{eq:max.comp:135}) the parameter $1/\eta$  determines the water-level, which decides whether a \ac{UE} is served or not, while $\beta_k$ is the cost
(in terms of complexity) of transmitting. If the difference between the selected rate $r_k$ and the capacity $\log(1 + \gamma_k)$
is small, then the term $\beta_k$ becomes very large (due to the large slope of $a_k$). Hence, the \ac{UE} is unlikely to be served.
The parameter $\alpha_k$ scales the rate, i.\,e. if again the rate $r_k$ operates close to capacity, $\alpha_k$ will be also very large and therefore scales down the assigned data rate in order to reduce the necessary computational complexity.

\subsection{Application to Limited Number of Rates and Multiple Users} \label{sec:Optimal_Scheduler}

The solution of (\ref{eq:max.comp:10}) is not practical in the more realistic case that the allocated rate must be drawn from among a finite set of discrete \ac{MCS} levels.  Assume now that the allocation is over the set $\Set{R}^* = \left\{\Set{R}_1, \dots \Set{R}_{N_A}\right\}$ of feasible discrete allocations.
The discretized allocation may be stated as:
\begin{eqnarray}
	r_k & \stackrel{(a)}{=} & \frac{1}{2\alpha_k}\left(\frac{1}{\eta} - \beta_k\right) \label{eq:practical.relevance:102} \\
& \stackrel{(b)}{\geq}  & \sqrt{\frac{C_k}{\alpha_k} + \left(\frac{\beta}{2\alpha_k}\right)^2} - \frac{\beta}{2\alpha_k} \label{eq:practical.relevance:104}
\end{eqnarray}
where $(a)$ follows from Theorem \ref{theorem:max_rate_scheduling} assuming that $0 \leq \eta \leq 1/\beta_k$
and $(b)$ results from (\ref{eq:max.comp:130}), assuming there is no constraint on the computational complexity. Note that in (\ref{eq:practical.relevance:104}) and in the following equations, $C_k$ is evaluated by (\ref{eq:max.comp:100}).


Combining (\ref{eq:practical.relevance:102}) and (\ref{eq:practical.relevance:104}), we can further state that
\begin{eqnarray}
	\frac{1}{\eta} & \geq &
2\alpha_k\sqrt{\frac{C_k}{\alpha_k} + \left(\frac{\beta}{2\alpha_k}\right)^2} \label{eq:practical.relevance:106} \\
	& = &
\sqrt{4\alpha_k C_k + \beta_k^2}, \label{eq:practical.relevance:108}
\end{eqnarray}
which gives us the required water-level for each user.

Equations (\ref{eq:practical.relevance:108}) and (\ref{eq:max.comp:342}) ($\eta > 1/\beta_k \implies  r_k = 0$) yield
\begin{equation}
	\sqrt{4\alpha_k C_k + \beta_k^2} \geq \beta_k \implies r_k = 0, \label{eq:practical.relevance:110}
\end{equation}
which is a condition that makes sure that the computational resources released are invested to those users, which can be served with higher assignable rates.

Another drawback of (\ref{eq:max.comp:10}) is that it does not assign the computational resources in a fair manner. In particular, the aforementioned solution always favors those users that have a higher \ac{SINR} and it allows them to transmit at the highest rates, as can be seen from function $l(\cdot, \cdot)$, while the other users might be dropped.


In order to schedule the rates more fairly from the set of allowed values, the following iterative procedure can be used:
\begin{enumerate}
      \item Initialization:
       	\begin{enumerate}
	  \item Set $\Set{R}$ such that each user $k$ receives the maximum possible rate $r_k$.
      \item Set $r_k=0$ for all users for which (\ref{eq:practical.relevance:110}) is satisfied.
	\end{enumerate}
      \item Recursion: If $\sum\limits_{r_k\in\Set{R}} \Set{C}_k > \Set{C}_\text{server}$ then \label{step:Recursion}
	\begin{enumerate}
	  \item Compute (\ref{eq:practical.relevance:108}) for each user and select $k^*$ user with the highest value: $k^* = \argmax_k  \sqrt{4\alpha C_k + \beta_k^2} $.
	  \item Decrease the rate for user $k$ to the next lower \ac{MCS} and update $\Set{R}$ accordingly.
	\end{enumerate}
      \item Decision:
        \begin{enumerate}
        \item  Halt the process if $\sum\limits_{r_k\in\Set{R}}\Set{C}_k \leq \Set{C}_\text{server}$. Otherwise, go back to step \ref{step:Recursion}.	 
        \end{enumerate}
      \item Attempt to serve dropped users: \label{step:Additional}
        \begin{enumerate}
             \item  Find the users for which $r_{k}=0$ and among them find  $k^*$ such that $ k^* = \argmax_k  \sqrt{4\alpha C_k + \beta_k^2}$. \label{step:Init}
             \item  Assign to the user $k^*$ the rate provided in 1.a and update $\Set{R}$ accordingly.
             \item  Halt the process if $\sum\limits_{r_k\in\Set{R}}\Set{C}_k  \geq \Set{C}_\text{server}$, otherwise go back to \ref{step:Init}. If $\sum\limits_{r_k\in\Set{R}}\Set{C}_k  > \Set{C}_\text{server}$ set $r_{k^*}=0$, and update $\Set{R}$ accordingly.
        \end{enumerate}
\end{enumerate}

In the previously described method, step 1-3 iteratively reduce the \ac{MCS} for the user with the highest value of (\ref{eq:practical.relevance:108}), until the complexity constraint is satisfied. Step \ref{step:Additional} attempts to assign eventual remaining computational resources to those users that were dropped even though their \ac{MCS} were sufficiently high. In the following, we refer to this process as \emph{\ac{SWF}}.




\subsection{Complexity Cut-Off}\label{sec:SCC_Scheduler}

In this section, we introduce a slightly simpler and intuitive scheduling method. Instead of applying (\ref{eq:max.comp:135}), which requires the linearization described in (\ref{eq:max.comp:20}) in order to determine the parameters $\alpha_k$ and $\beta_k$, we select the users with the highest complexity and reduce their \ac{MCS} (and associated rate $r_k$) until the sum-complexity constraint $\Set{C}_\text{server}$ is fulfilled. In this case an iterative method, similar to one described in Sec. \ref{sec:Optimal_Scheduler} can be used, which works as follows:
\begin{enumerate}
	\item Initialization: Set $\Set{R}$ such that each user $k$ receives the maximum possible rate $r_k$,
	\item Recursion: If $\sum\limits_{r_k\in\Set{R}} \Set{C}_k > \Set{C}_\text{server}$ then
	\begin{enumerate}
		\item Select the user $k^*$ with the highest complexity: $k^* = \argmax_k \Set{C}_k $
		\item Decrease the rate for user $k$ to the next lower \ac{MCS} and update $\Set{R}$ accordingly
	\end{enumerate}
	\item Decision: Halt the process if $\sum\limits_{r_k\in\Set{R}}\Set{C}_k \leq \Set{C}_\text{server}$. Otherwise, go back to step 2.	
\end{enumerate}

As the result of this procedure, we obtain a rate allocation $\Set{R}\in\Set{R}^*$, which satisfies the complexity constraint and always reduces the rate of those users  that require the higher complexity. The main difference to the previously discussed allocation algorithm is that this is not necessarily sum-rate optimal. However, it is still very efficient as we will discuss in the next section. In the following, we refer to this method as \emph{\ac{SCC}}.

%
%
%
%
%

\section{System-level Evaluation}\label{sec:results}
In this section, we evaluate the performance of the proposed resource allocation strategies in terms of complexity and sum-rate. The performance of \ac{SWF}, provided in Sec. \ref{sec:Optimal_Scheduler}, is compared against \ac{SCC}, described in Sec. \ref{sec:SCC_Scheduler}.  In addition, the schedulers are compared against the benchmark \ac{MRS} policy \cite{Valenti.Talarico.Rost.GC.2014}, which is the scheduler that simply sets each $r_k$ to its maximum value by selecting the maximimum rate that satisfies the outage constraint after $L_\mathsf{max}$ decoder iterations for the given SINR $\gamma_k$.  The comparisons are made by using a system-level simulator that is compliant with the 3GPP LTE standard.

\subsection{System-Level Simulator}
We assume a 3GPP LTE system using adaptive modulation and coding based on turbo codes with overall 27 distinct \acp{MCS} ($N_R=27$). The rate of the $i^{th}$ \ac{MCS} is given by \cite{Rost.Talarico.Valenti.TWC.2015}
\begin{eqnarray}
  r_i
  & = &
  \log_2\left( 1+ \frac{\gamma^{R}_{i}}{\nu } \right ),
\end{eqnarray}
where $\gamma^R_{i}$ indicates the minimum \ac{SINR} for which the $i^{th}$ \ac{MCS} satisfies on average an outage constraint after the $L_\mathsf{max}$-th iteration, while $\nu$ is a parameter that models the gap between the capacity at $\gamma^{R}_{i}$ and the \ac{SINR} for the actual code to meet the performance objective at rate $r_i$.
In the following, it is assumed that $L_\mathsf{max}=8$. The value of $\gamma^R_{i}$ for each \ac{MCS} can be obtained as follows.  Simulations are used to obtain \ac{TBLER} curves for each possible \ac{MCS} by setting an upper bound on the maximum number of turbo-iterations. For the $i^{th}$ \ac{MCS}, $\gamma^R_{i}$ is selected to be the value of \ac{SINR} for which the \ac{TBLER} satisfies a particular constraint for the channel outage $\hat\epsilon_\text{channel}$.

The parameter $\nu$ together with the complexity model parameters $\{K',\zeta\}$ are selected by statistically fitting the model with an actual LTE turbo-decoder. In particular, from \cite{Rost.Talarico.Valenti.TWC.2015} the best fit is given for $K'=0.2$, $\zeta=6$, and $\nu=0.2$ dB.

\begin{figure}
      \centering
      \includegraphics[width=7cm]{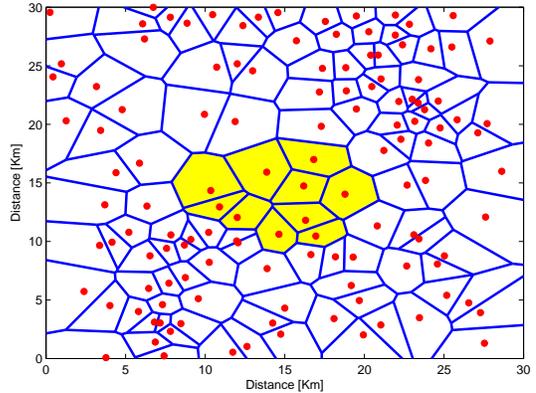}
      \caption{\ac{BS} locations with centrally processed \acp{RAP} highlighted in yellow.}
      \label{fig:cellular_network}
\end{figure}

We consider a network composed of $N_\mathsf{bs}=129$ \acp{BS} shown in Fig. \ref{fig:cellular_network}, which is a segment of the actual deployment by a major provider in the UK at $1800$ MHz over a square arena of $30\times30$ $\text{km}^2$. Assume that the Cloud-\ac{RAN} platform centrally processes the uplink signals from the $N_\mathsf{c}=10$ cells highlighted in yellow and the \acp{UE} are distributed according to a \ac{PPP} with intensity $\lambda$ users per $\mathsf{km}^2$.


Let $Y_k$ indicate the $k^{th}$ \ac{RAP} that serves a cell with area $\mathcal A_k$ and its location, and $X_k$ indicate the \ac{UE} served by the $k^{th}$ \ac{RAP} and its location.
We assume the path loss from a mobile $X_k$ to a \ac{BS} $Y_k$ is  $|Y-X|^{-\alpha}$, where $\alpha$ is the path-loss exponent.  When fractional power control is used, a mobile's transmit power is $P_k = P_0 |Y_k - X_k|^{s \alpha}$, where $P_0$ is the reference power (which is assumed as the power received at unit distance from the transmitter) and $s, 0 \leq s \leq 1,$ is the compensation factor for fractional power control. In the following, we assume $s=0.1$, which is the value reported in  \cite{coupechoux:2011} that maximizes the sum throughput.

The fading power gain from $X_i$ to $Y_k$ is assumed to be exponentially distributed with unit mean, corresponding to Rayleigh fading. Furthermore, assume that the fading power gains remain fixed for the duration of a transmission, but vary from user to user and from one \ac{TB} to the next (block fading).


\begin{table}
 \caption{Main parameters for the system level evaluation.}
    \centering
    \small
      \begin{tabular}{|p{5.0cm}||p{2.5cm}|}
      \hline
	Spatial distribution of users & PPP \\ \hline
	Density of \ac{UE} per unit area & $\lambda=1$ UEs/$\mathsf{Km}^2$ \\ \hline
	Path loss exponent  & $\alpha= 3.7$ \\ \hline
	Number of centrally processed RAPs & $N_\text{c}= 10$ \\ \hline
	Computational outage  & $\epsilon =\left[10, 1, 0.1\right]{\%}$ \\ \hline
Channel outage constraint & $\hat \epsilon_\text{channel}=\unit[10]{\%}$ \\ \hline
	Fading        &   Rayleigh           \\ \hline
	Fractional Power-Control Factor     &   s=0.1\\ \hline
	Transmit power    &   $P_0=10$ W \\ \hline
	Noise power    &   $W=100$ mW \\ \hline

	Simulation trials       &   $N_\text{trials}= 10^7$        \\ \hline
      \end{tabular}
\label{main_table}
\end{table}

The sum-rate and the sum-complexity of the system are used in the following as performance metrics to compare the proposed allocation strategies with the benchmark scheduler. Both performance metrics are evaluated through simulations for each of the allocation strategies under examination as follows. During each trial, a mobile is placed at random in the $k^{th}$ cell with probability $1-\exp( - \lambda \mathcal A_k )$.  Once the mobiles are placed, the fading coefficients are drawn from an exponential distribution and the \ac{SINR} at each of the $N_c$ \acp{RAP} is computed. By applying a given allocation strategy, we find the selected \ac{MCS} for each \ac{RAP} and the corresponding rate based on the quality of the channel. Using (\ref{eq:max.comp:100}), the complexity required to process the uplink signal of each \ac{UE} is evaluated. The sum-rate and sum-complexity are now computed by summing up respectively the rates and complexity for all $N_c$ \acp{RAP}. Once the allocation algorithm is applied, if $\sum_{k}\Set{C}_k > \Set{C}_\text{server}$, a computational outage occurs and the sum-rate is set to zero.

\subsection{Numerical Results}

In this subsection, the parameters summarized in Table \ref{main_table} are used in the system-level simulator, if not otherwise stated.

\begin{figure}
	\centering
	\subfigure[CDF of the computational complexity.]{\includegraphics[width=6.5cm]{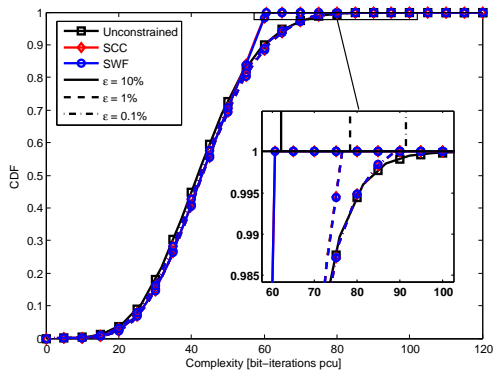}}
	\subfigure[CDF of the sum-rate.]{\includegraphics[width=6.5cm]{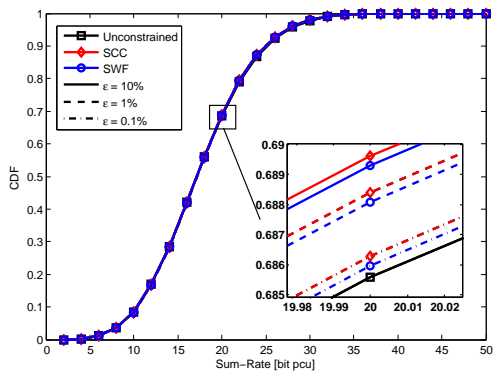}}
	\caption{CDFs of the complexity and of the sum-rate. 
	}
	\label{fig:results:cdf_complexity_sum_rate}
\end{figure}

Fig. \ref{fig:results:cdf_complexity_sum_rate} shows the \ac{CDF} of both the achieved sum-rate and the required computational complexity over $N_c = 10$ \acp{RAP}. The figure shows the curves for both \ac{SWF} and \ac{SCC}, when the system is designed such that $\epsilon=\unit[10]{\%}, \unit[1]{\%}, \unit[0.1]{\%}$ computational outage holds (the notches in the magnification of Fig. \ref{fig:results:cdf_complexity_sum_rate}(a) show the corresponding value of $\Set{C}_\text{server}$). Each of these values translates into a peak-requirement for computing resources given the respective outage constraint. Fig. \ref{fig:results:cdf_complexity_sum_rate} shows as a benchmark the curve for the unconstrained case, which selects the maximum possible rate that can be used since there is no computational constraint. Fig. \ref{fig:results:cdf_complexity_sum_rate}(a) shows that a stronger constraint on the available computational resources implies higher computational outage. More importantly, this figure  highlights that while the required computational complexity is significantly reduced by dimensioning the system to an higher computational outage, the sum-rate only decreases slightly for both \ac{SWF} and \ac{SCC}, i.\,e. the average sum-rate only decreases by $\approx\unit[0.28]{\%}$ for $\epsilon = \unit[10]{\%}$ and by $\approx \unit[0.07]{\%}$ for $\epsilon = \unit[0.1]{\%}$, as illustrated by Fig. \ref{fig:results:cdf_complexity_sum_rate}(b).

This shows the efficiency of the proposed schedulers, which impact the achievable sum-rate only marginally, while they reduce the required computational resources significantly. Furthermore, Fig. \ref{fig:results:cdf_complexity_sum_rate} shows that the introduced schedulers are able to completely avoid any computational outage, which would lead in the worst case to drop the connection of \acp{UE}. Even if one solution could be to dimension the system for a very low computational outage, i.\,e. $\epsilon = \unit[10^{-6}]{\%}$, the drawback is that the system will be significantly over-provisioned and most of the time the allocated resources are under-utilized. By contrast, our schedulers allow to avoid computational outage, while maintaining a high server utilization.

\begin{figure}
	\centering
	\includegraphics[width=6.5cm]{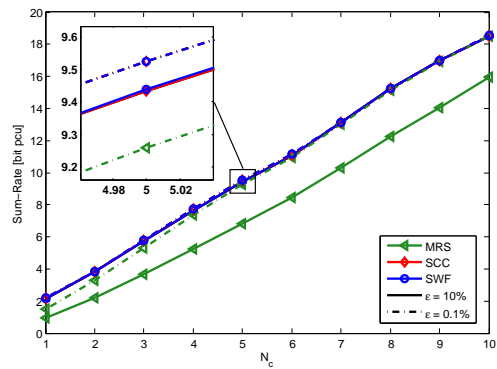}
	\caption{Sum-rate as a function of $N_c$.
	}
	\label{fig:results:sumrate_over_N}
\end{figure}

\begin{figure}
	\centering
	\includegraphics[width=6.5cm]{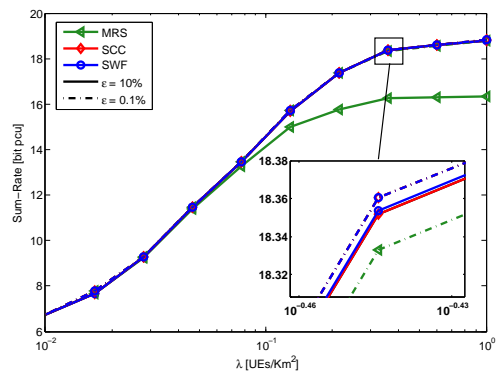}
	\caption{Sum-rate as a function of the user-density $\lambda$.
	}
	\label{fig:results:sumrate_over_lambda}
\vspace{-0.20cm}
\end{figure}

Fig. \ref{fig:results:sumrate_over_N} shows the sum-rate as a function of the number of \acp{RAP}, which are centrally processed at the  Cloud-\ac{RAN} platform. The performance figures are shown for \ac{SWF} and \ac{SCC} as well as for \ac{MRS}, which does not account for the computational constraint. Fig. \ref{fig:results:sumrate_over_N} shows the impact of the computational outage targets on the different schedulers. As it can be noticed, for all values of $N_c$, the impact of the constraint on the computational resources is marginal for the computational aware schedulers. By contrast, as $\epsilon$ increases, the impact on a system, which uses \ac{MRS}, increases linearly with $\epsilon$ due to the increasing computational outage. Furthermore, the magnification in Fig. \ref{fig:results:sumrate_over_N} shows that there is only a marginal difference between \ac{SWF} and the \ac{SCC}, emphasizing the fact that the less complex \ac{SCC} algorithm achieves almost the same performance as \ac{SWF}.

In mobile networks, it may happen that the traffic demand differs significantly over time. In this case the system experiences traffic peaks, e.g. when many people leave or join the metro or an event. This may lead to cases where the system experiences a higher computational demand than usual. Fig.\ \ref{fig:results:sumrate_over_lambda} shows the sum-rate of \ac{SWF}, \ac{SCC} and \ac{MRS} as function of the density of \acp{UE} per $\text{km}^{2}$ when the maximum computational resources available lead to a computational outage of $\unit[10]{\%}$ and $\unit[0.1]{\%}$ in the case of $\lambda = \unit[0.5]{\text{UEs/km}^2}$. Fig. \ref{fig:results:sumrate_over_lambda} shows the ability of our proposed schedulers to provide services to all the users while slightly reducing the system-throughput. Furthermore, it is shown that the proposed scheduling algorithms are able to accommodate the increasing traffic demand, while the \ac{MRS} suffers gradually from the higher computational outage.

\vspace{-0.20cm}

\section{Conclusions}\label{sec:conclusion}
The computational complexity of \ac{RAN} functions is one of the main obstacles for the introduction of cloud computing principles into the mobile network radio access. In this paper, we have developed a framework, which solves the user resource allocation problem under the assumption of limited computational resources in a centralized cloud platform. We showed that the underlying optimization problem can be solved with an adapted water-filling approach, making it feasible to fulfill the strict timing requirements of the wireless access (e.g., several milliseconds in LTE). Furthermore, we have shown that an intuitive complexity-cut-off approach delivers near-to-optimal results as well.
Finally, the numerical evaluation confirms that meeting computational complexity constraints does not lead to significant penalties in terms of throughput, a fact which underlines the applicability of the approach in practical systems.


\vspace{-0.20cm}
\appendices
\section{Proof of Theorem \ref{theorem:max_rate_scheduling}}\label{sec:proof.theorem:max_rate_scheduling}
  \begin{IEEEproof}
  This section provides details leading to the solution of the optimization problem given by (\ref{eq:max.comp:10}). Since both (\ref{eq:max.comp:130}) and the constrained functions in (\ref{eq:max.comp:10}) have continuous first partial derivatives, this problem can be solved through the method of Lagrange multipliers. Given the Lagrange multipliers $\eta$ and $\mathbf{\Theta}=\{ \Theta_1,...,\Theta_{N_c}\}$, the Lagrangian can be written as follows
    \begin{align}
      L(\Set{R}, \eta, \hspace{-0.05cm} \mathbf{\Theta}) = - \hspace{-0.15cm} \sum\limits_{r_k\in\Set{R}} \hspace{-0.1cm} r_k
	 \hspace{-0.05cm} + \eta\hspace{-0.05cm}\left(\sum\limits_{r_k\in\Set{R}} \hspace{-0.1cm} \mathcal{C}_k  \hspace{-0.05cm} -  \hspace{-0.05cm} \mathcal{C}_\text{server}\right)
   \hspace{-0.05cm} - \hspace{-0.05cm} \Tr\left[\mathbf{\Theta}\diag(r_k)\right]. \nonumber
    \end{align}
    The partial derivative of the Lagrangian over $r_k$ is
    \begin{eqnarray}
      \frac{\partial L}{\partial r_k} & = & -1 + \eta\frac{\partial \mathcal{C}_k}{\partial r_k} - \mathbf{\Theta} \nonumber \\
       & = & -1 + \eta\left(2\alpha_k r_k + \beta_k\right) - \mathbf{\Theta}.
       \label{eq:partial_derivative}
    \end{eqnarray}
    Using the Karush-Kuhn-Tucker conditions, it follows that
   \begin{eqnarray}
	\hspace{-0.5 cm} \forall k: \frac{\partial L}{\partial r_k} =0  \hspace{-0.35 cm} &\implies & \hspace{-0.35 cm}
	  1 + \Theta_k  = \eta\left( 2\alpha r_k + \beta_k\right) \label{eq:First_condition}
\end{eqnarray}
 \begin{eqnarray}
	\hspace{-0.5 cm} \sum\limits_{r_k\in\Set{R}} \mathcal{C}_k \leq \mathcal{C}_\text{server} \hspace{-0.35 cm} &\implies& \hspace{-0.35 cm}
	\eta \hspace{-0.05 cm} \geq \hspace{-0.05 cm} 0 \label{eq:Second_condition}
      \\
	\hspace{-0.5 cm} \forall k \hspace{-0.05 cm} : \hspace{-0.05 cm} r_k  \hspace{-0.05 cm} \geq \hspace{-0.05 cm} 0 \hspace{-0.35 cm} &\implies& \hspace{-0.35 cm}
	\forall \Theta_k \hspace{-0.05 cm} \geq  0 :
  \Theta_k r_k =  0.  \label{eq:Third_condition}
\end{eqnarray}

    First, lets assume that $r_k\neq0\implies\Theta_k=0$, which follows from (\ref{eq:Third_condition}). Using (\ref{eq:First_condition}) and setting $\Theta_k=0$, it yields
    \begin{eqnarray}
      1 & = & \eta\left(2\alpha_k r_k + \beta_k\right).
      \label{eq:NonZeroRate}
    \end{eqnarray}
    From (\ref{eq:NonZeroRate}) using (\ref{eq:Second_condition}), it follows that (with $\eta\geq0$)
    \begin{eqnarray}
      r_k & = & \frac{1}{2\alpha_k}\left(\frac{1}{\eta} - \beta_k\right)^+.\label{eq:max.comp:340}
    \end{eqnarray}

    Lets consider the case when $\Theta_k\neq0\implies r_k=0$, which again follows from (\ref{eq:Third_condition}). Using (\ref{eq:First_condition}) and $r_k=0$, it yields
    \begin{eqnarray}
      1 + \Theta_k & = & \eta\beta_k.
      \label{eq:NonZeroTheta}
    \end{eqnarray}
    From (\ref{eq:NonZeroTheta}) and since in this case $\Theta_k>0$,  it follows that
        \begin{eqnarray}
      \eta > \frac{1}{\beta_k}.
      \label{eq:max.comp:342}
    \end{eqnarray}
    By combining (\ref{eq:max.comp:340}) and (\ref{eq:max.comp:342}), Theorem \ref{theorem:max_rate_scheduling} is obtained.
  \end{IEEEproof}
  
\vspace{-0.20cm}

\bibliographystyle{IEEEtran}
\bibliography{IEEEfull,references}

\balance
\vfill

\end{document}